# Potential Vorticity of
# the South Polar Vortex of Venus

**I. Garate-Lopez (1), R. Hueso (1), A. Sánchez-Lavega (1), A. García Muñoz (2, 3)**

(1) Departamento de Física Aplicada I, E.T.S. Ingeniería, Universidad del País Vasco, Alameda Urquijo s/n, 48013 Bilbao, Spain.

(2) ESA Fellow, ESA/RSSD, ESTEC, 2200 AG Noordwijk, The Netherlands.

(3) Zentrum für Astronomie und Astrophysik, Technische Universität Berlin, 10623 Berlin, Germany.

**Key Points**

- We present Ertel's potential vorticity (EPV) maps of Venus' south polar vortex.
- The relative vorticity dominates over the thermal term in the EPV calculation.
- EPV is ~$2\times10^{-6}$ and ~$2\times10^{-8}$ $K\cdot m^2\cdot kg^{-1}\cdot s^{-1}$ in the upper and lower cloud respectively.

Manuscript pages: 41

Figures: 11

Tables: 0

Corresponding author address:

Itziar Garate Lopez

Dpto. Física Aplicada I,

E.T.S. Ingeniería, Universidad del País Vasco,

Alda. Urquijo s/n, 48013 Bilbao, Spain

**E-mail:** itziar.garate@ehu.eus

**Telephone:** (34) 94 601 7389

**Fax:** (34) 94 601 4178

**Abstract**

Venus' atmosphere shows highly variable warm vortices over both of the planet's poles. The nature of the mechanism behind their formation and properties is still unknown. Potential vorticity is a conserved quantity when advective processes dominate over friction and diabatic heating, and is a quantity frequently used to model balanced flows. As a step toward understanding the vortices' dynamics, we present maps of Ertel's potential vorticity (EPV) at Venus' south polar region. We analyze three configurations of the South Polar Vortex at the upper cloud level (P~240 mbar; z~58 km), based on our previous analyses of cloud motions and thermal structure from data acquired by the VIRTIS instrument onboard Venus Express. Additionally, we tentatively estimate EPV at the lower cloud level (P~2200 mbar; z~43 km), based on our previous wind measurements and on static stability data from Pioneer Venus and the VIRA model. Values of EPV are on the order of $10^{-6}$ and $10^{-8}$ K·m$^2$·kg$^{-1}$·s$^{-1}$ at the upper and lower cloud levels, respectively, being 3 times larger than the estimated errors. The morphology observed in EPV maps is mainly determined by the structures of the vertical component of the relative vorticity. This is in contrast to the vortex's morphology observed in 3.8 or 5 µm images which are related to the thermal structure of the atmosphere at the cloud top. Some of the EPV maps point to a weak ringed structure in the upper cloud while a more homogenous EPV field is found in the lower cloud.

# 1. Introduction

During the nominal and part of its extended mission (from April 2006 to October 2008), the Venus Express spacecraft obtained detailed images of the thermal structure in the South polar region of the planet, which behaves as an atmospheric warm vortex at cloud top [Piccioni et al., 2007; Titov et al., 2012; Garate-Lopez et al., 2015]. The South Polar Vortex (SPV) is similar to the circumpolar vortex of the North Hemisphere observed by previous missions at visual wavelengths [Suomi and Limaye, 1978] and in the thermal infrared [Taylor et al., 1979, 1980; Schofield and Diner, 1983]. The SPV cloud morphology, and its temporal evolution and lifetime have been investigated with data from the VIRTIS (Visual and InfraRed Thermal Imaging Spectrometer) and VMC (Venus Monitoring Camera) instruments. Luz et al. [2011] and Garate-Lopez et al. [2013] presented detailed accounts of the vortex's morphology and its cloud motions. The three-dimensional thermal structure of the vortex was first investigated by Grassi et al. [2008] and has also been the subject of a recent detailed study based on VIRTIS data for three specific vortex configurations [Garate-Lopez et al., 2015].

The combination of velocity and thermal structure determines the behavior of atmospheric structures through its combination into potential vorticity in one of its many different formulations [Pedlosky, 1987; Sánchez-Lavega, 2011]. Ertel's potential vorticity (EPV) is a key atmospheric variable used in diagnostic and prognostic models of geophysical fluids because, for an inviscid atmospheric flow (neglecting friction) and in the absence of sinks or sources of potential vorticity from diabatic heating, EPV is a conserved quantity and becomes a tracer of fluid motions [Sánchez-Lavega, 2011]. In fact, any dynamical field in a fluid can be determined given the global distribution of isentropic potential vorticity (e.g. the EPV), the mass under each isentropic surface and appropriate boundary conditions [Hoskins et al., 1985; Vallis, 2006].

On Earth, potential vorticity maps are a tool commonly used to study the evolution of the stratospheric polar vortices [Nash et al., 1996]. The jet stream and the cyclonic circulation around terrestrial polar vortices act as a barrier to mixing, and are responsible for the intensity of the ozone hole [Shoeberl et al., 1992]. Relatively low values of total column ozone and cold temperatures in the lower stratosphere are found co-located with polar vortices [Schoeberl and Hartmann, 1991]. Both polar vortices are strongly seasonally dependent, but the winter southern polar vortex is larger, more intense and longer-lasting than its northern counterpart.

Latitudinal confinement by a mixing barrier could be important in polar vortices present in other Solar System planets. Mitchell et al. [2014] used EPV to compare the polar vortices on Earth and Mars. Unlike on Earth, the Martian polar vortices are annular and become dramatically smaller with height. Teanby et al. [2008] compared EPV maps with measurements of five independent chemical tracers to study Titan's winter polar vortex and found that indeed the vortex circumpolar jet separates a tracer-enriched air mass in the North Pole from air at lower latitudes. Permanent, strong polar vortices confined by narrow jet streams exist also at the poles of Saturn as observed in the cloud [Sánchez-Lavega et al., 2006; Dyudina et al., 2008; Antuñano et al., 2015] and temperature fields [Fletcher et al., 2008]. Therefore a comparative view of the dynamics of these vortices in very different environments could help to understand the mechanisms behind their formation and temporal variability.

The Venus atmosphere shows an enhancement of CO, a trace gas in Venus atmosphere, from the equator to the south pole with a peak at ~60°S and 35 km altitude [Tsang et al., 2008]. The decrease of CO from 60°S to the pole could be an evidence of the existence of a latitudinal mixing barrier with the CO enhancement at 35 km caused by the descending branch of a Hadley cell that may advect CO from the cloud top, where CO is produced by photodissociation, to this lower altitude [Tsang et al., 2008]. Piccialli et al. [2012] investigated this possibility by calculating the zonal mean of EPV from thermal vertical profiles at subpolar latitudes obtained by the VeRa instrument. This data was used to compute zonal mean thermal winds and EPV. The values of EPV from this study slightly increase from equator to pole but they do not show any mixing barrier or region of strong latitudinal gradient. However, they only studied the zonal mean of the EPV and assumed cyclostrophic balance which constitutes a physical approximation that fails in reproducing the winds obtained by cloud tracking close to the equator and the poles.

Here, we construct horizontal maps of the EPV field at the upper cloud level (about 58 km above the surface) over the south polar region of Venus. For that purpose we combine simultaneously obtained maps of wind and temperature that we derived previously from observations acquired by the VIRTIS instrument on board Venus Express [Garate-Lopez et al., 2013; Garate-Lopez et al., 2015]. Additionally, we tentatively estimate the EPV distribution at the lower cloud level (about 43 km altitude) combining our previous wind measurements [Garate-Lopez et al., 2013] and combined values of the static stability from Pioneer Venus North probe, Pioneer Venus radio-occultation experiment and the VIRA model [Seiff et al., 1980, 1985] which we consider as representative of the possible thermal structure of the lower cloud level in the south polar atmosphere. We examine three different configurations of the vortex in order to comprehend the relation of the vortex morphology with its dynamical properties and its long-term behavior. Figure 1 shows images of the vortex's upper and lower clouds in the three configurations. The hourly evolution of the potential vorticity in the upper cloud is also investigated for the last of these configurations, which was observed during orbit 475 when VIRTIS obtained high-resolution observations of the vortex every 15 minutes over several hours.

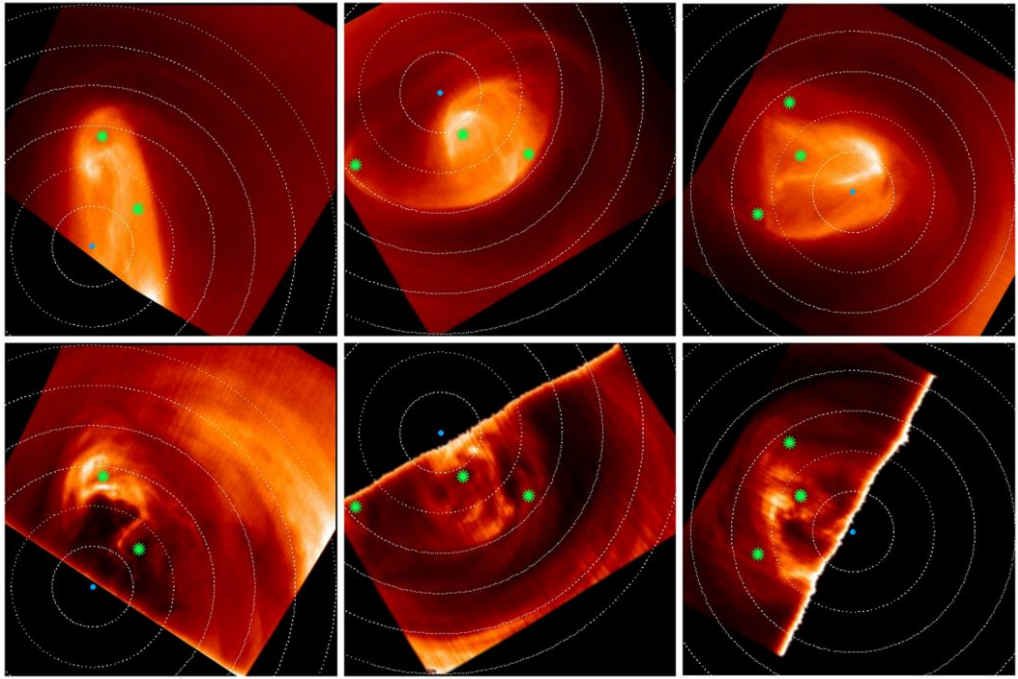

**Figure 1:** Polar projected images showing the morphology of the South Polar Vortex of Venus at the upper (top) and lower (bottom) cloud levels on orbits 038 (left), 310 (center) and 475 (right). The upper cloud is observed at 3.8 or 5.1 µm VIRTIS-M-IR images and the lower cloud at 1.74 µm images. Latitude circles are plotted at 5° intervals from the south pole (represented by a blue dot). Green stars have been added to stress the vertical consistency of the overall shape of the vortex.

Deriving EPV fields in the polar region of Venus's atmosphere has been significantly challenging due to the limitations of the actual datasets (mainly horizontal and vertical spatial resolutions) and the highly variable nature of the vortex. However, they are likely to be the only available EPV estimates for some time to come, since the Japanese Akatsuki mission will not be able to observe the polar latitudes from its equatorial orbit.

Section 2 summarizes the wind and temperature analyses. Section 3 describes the calculation of Ertel's potential vorticity. We present the results in Section 4, and in Section 5 we discuss the results and present our conclusions.

## 2. Wind Measurement and Temperature Retrieval

### 2.1. Wind field analysis

We measured wind motions at the southern pole of Venus [Garate-Lopez et al., 2013], from the analysis of images obtained with the infrared channel of the VIRTIS-M instrument over the 1.0 – 5.1 µm spectral range [Drossart et al., 2007]. The data consists of imaging qubes with two spatial dimensions of up to 256 x 256 pixels and one spectral dimension. Night-side images at 1.74 µm are sensitive to the structure of the lower cloud (at about 42 km altitude above the surface at polar latitudes [Barstow et al., 2012]) and constitute the main data to study lower cloud dynamics [Sánchez-Lavega et al., 2008; Hueso et al., 2012; Hueso et al., 2015]. Images at 3.8 and 5.1 µm are sensitive to the thermal emission of the upper cloud (at about 63 km altitude at polar latitudes [Ignatiev et al., 2009; Peralta et al., 2012]). In some cases the characteristics of the VIRTIS hyperspectral images allowed to retrieve the simultaneous wind field at both levels. This was done by analyzing images obtained when the spacecraft was close to the apocenter in near-nadir pointing, achieving a spatial resolution of ~16 km x 16 km for each individual pixel. Feature motions were extracted with an image correlation algorithm [Hueso et al., 2009] that allows to manually filter spurious measurements on the fly. The selected image-pairs were separated in time by $1 - 2$ hours and uncertainties in each individual wind measurement were estimated to be about 4 m/s. We here revise this figure and increase it to 6 m/s to take into account the larger uncertainty of some tracked features.

Maps of the vertical component of the relative vorticity:

$$\zeta(\lambda, \varphi) = \frac{1}{R\cos\varphi}\frac{\partial v}{\partial \lambda} - \frac{1}{R}\frac{\partial u}{\partial \varphi} + \frac{u}{R}\tan\varphi \, , \tag{1}$$

with u and v being the zonal and meridional wind velocities, $\varphi$ the latitude, $\lambda$ the longitude and R the radius of the planet, were obtained from the wind measurements for different vortex configurations (see Figure 2 in Garate-Lopez et al. [2013]). These derivatives were calculated using a coordinate transformation into a rectangular x-y grid centered in the pole and spatial derivatives were calculated with a spatial step of 5° (525 km) to reduce errors associated with small irregularities in the wind field. In both cloud levels, the polar vortex is a relatively weak vortex immersed in a cyclonic environment whose cloud and thermal morphology is not directly related to the structures observed in the relative vorticity maps. Higher mean vorticity values about $\zeta \sim (6.0 \pm 3.5) \times 10^{-5}\,\mathrm{s}^{-1}$ were found at polar latitudes (75°S – 90°S) when compared with sub-polar latitudes (60°S – 75°S) where $\zeta \sim (2.5 \pm 3.5) \times 10^{-5}\,\mathrm{s}^{-1}$. The uncertainties here reflect the increased uncertainty in wind measurements with respect to Garate-Lopez et al. [2013].

## 2.2. Temperature field analysis

VIRTIS infrared spectra from $4.2 - 5.1$ µm contain enough information to retrieve temperatures from about 55 to 85 km altitudes [García Muñoz et al., 2013]. We used an inversion relaxation technique that tries to find a best match between an observed spectrum and a modeled spectrum by iteratively correcting an initial guess temperature profile to a final thermal profile. We used synthetic spectra generated from an atmospheric model as described in García Muñoz et al. [2013]. The retrieval algorithm largely followed the methodology by Grassi et al. [2008] with changes in the way clouds and aerosols are treated. Instead of considering the cloud top altitude and the aerosol scale height as free parameters within the iteration algorithm, we retrieved temperatures for a discrete set of fixed values of both parameters and then selected the cloud parameters and thermal profile that sowed a better fit of the modeled and observed spectra (in terms of the root-mean-square deviation). The thermal retrievals are fully explained in Garate-Lopez et al. [2015].

We selected three of the orbits where we measured wind fields with enough spatial resolution and provided high-quality spectral data over the $3-5$ µm range in most of the image area. These three cases show very different morphologies of the vortex (see Figure 1) and correspond to Venus Express orbits 038 (28 May 2006), 310 (24 February 2007), and 475 (8 August 2007). Thermal maps at altitudes from 55 km (~360 mbar pressure level) to 85 km (~1 mbar) are given in Figures 4 to 6 in Garate-Lopez et al. [2015].

In the Venusian atmosphere the temperature increases downwards from 100 to ~40 km, except for an inversion layer (at about $60-70$ km) coincident with the cold collar [Taylor et al., 1980; Seiff, 1983; Piccialli et al., 2008; Tellmann et al., 2009]. The cold collar is a ring of colder air that surrounds the vortex and is observed as a darker area between $60°$ and $80°$ in the upper panels of Figure 1. Our retrieved thermal profiles agree well with the mentioned trend and reproduce the inversion (a temperature decrease of up to 20 K in just 10 km altitude) at locations where the cold collar is observed. On orbit 038 the cold collar is more pronounced showing temperatures on the order of 220 K, while on orbit 475 the cold collar temperatures increase to 235 K. The latter case also shows the highest temperature values for the warm vortex [Garate-Lopez et al., 2015]. At ~68 km altitude the collar is on average 13 K colder than the mean vortex temperature but the difference increases up to 30 K as we go downwards in the atmosphere. These temperature differences between the warm vortex and the cold collar are much larger than the associated errors, which were estimated to be about 3 K on average at the whole altitude range ($55-85$ km) but increase up to 9 K in the lowest ~7 km.

## 3. Calculation of Ertel's Potential Vorticity

The general definition of Ertel's potential vorticity (EPV) under the hydrostatic approximation can be written as [Pedlosky, 1987; Sánchez-Lavega, 2011]:

$$q = \frac{\overrightarrow{\omega_R} + 2\overrightarrow{\Omega}}{\rho} \nabla\theta \sim (\zeta_\theta + f)\left(-g\frac{\partial\theta}{\partial P}\right), \qquad (2)$$

with $\overrightarrow{\omega_R} = \nabla \times \overrightarrow{U}$ being the vorticity of the wind vector $\overrightarrow{U}$, $\overrightarrow{\Omega}$ the angular rotation speed of the planet, $\rho$ the density, $\theta$ the potential temperature, $f = 2\Omega\sin\phi$ the Coriolis parameter (with $\phi$ being latitude), P the pressure and g the gravitational acceleration. On Venus, the Coriolis parameter f can be neglected, since it is two orders of magnitude lower than $\zeta_\theta$ (for example, at $\phi = 80°S$, $f \sim 6 \times 10^{-7}s^{-1}$ while $\zeta_\theta \sim 6 \times 10^{-5}s^{-1}$), so that equation (2) becomes:

$$q \sim \zeta_\theta\left(-g\frac{\partial\theta}{\partial P}\right), \qquad (3)$$

where the vertical component of the relative vorticity ($\zeta_\theta$) is calculated on an isentropic surface ($\theta = $ constant). This definition is valid for a vertically stable atmosphere, a condition that is globally fulfilled on Venus above 45 km according to the atmosphere's thermal structure found from VeRa data [Tellmann et al., 2009]. The topmost cloud layer and the atmosphere above are extremely stable to vertical motions. This high stability decreases below the upper clouds and the vertical lapse rate approaches to adiabatic at the middle cloud altitudes (50 – 55 km) being possible to find convective regions in shallow layers [Tellmann et al., 2009]. Therefore, equation (3) should not be used to derive EPV fields at ~50 – 55 km.

In our detailed thermal study of the vortex [Garate-Lopez et al., 2015] we calculated the static stability distribution at the upper cloud level and obtained values on the order of 8 – 14 K/km for the three vortex configurations under analysis. Accordingly, Piccialli et al. [2012], based on VeRa data, show large values of the Richardson number at vertical levels that are in agreement with the cloud top altimetry and its latitudinal profile independently derived from VMC and VIRTIS datasets [Ignatiev et al., 2009].

On the other hand, the static stability profiles from the Pioneer Venus radio-occultation experiment and VIRA model show another local maximum close to 43 km [Seiff et al., 1985] where the lower cloud is located in the polar regions. These vertical profiles display a decrease in the static stability towards the pole, also present on VeRa static stability profiles [Tellamnn et al., 2009]. Hence, we use equation (3) to tentatively estimate the EPV distribution at the lower cloud level, even though at the highest latitudes it may not be a good approximation since the static stability values are small.

In this paper, we will focus on the two regions of high static stability, the uppermost one located at the upper cloud level and the lower cloud located close to 45 km altitude where the static stability increases again [Tellmann et al., 2009].

### 3.1. Potential temperature for an atmosphere with temperature-dependent specific heat

The potential temperature $\theta$ is the temperature that an air parcel would have if it were moved adiabatically (no heating, cooling or mixing) from a level with temperature $T$ and pressure $P$ to a reference level with pressure $P_0$ [Sánchez-Lavega, 2011]:

$$\theta = T\left(\frac{P_0}{P}\right)^k,$$ (4)

where $k = \frac{(\gamma-1)}{\gamma} = \frac{R^*}{C_P}$ is the adiabatic index with $\gamma$ being the adiabatic coefficient, $R^*$ the specific gas constant ( $R^* = \frac{R}{M_{CO_2}} = 0.1889 \, \mathrm{J\,g^{-1}\,K^{-1}}$ , where $R = 8.3143 \, \mathrm{J\,mol^{-1}\,K^{-1}}$ and $M_{CO_2} = 44.01 \, \mathrm{g\,mol^{-1}}$), and $C_P$ the specific heat capacity at constant pressure.

This definition is obtained after considering the perfect gas law and assuming $C_P$ is a temperature independent constant. However, for Venus' atmosphere, where temperatures extend over a large range, it is necessary to consider an explicit dependence of $C_P$ on temperature. The specific heat at constant pressure of a gas essentially constituted by linear molecules such as $CO_2$ (96.5% on Venus atmosphere) can be approximated as a series of powers in T [Epele et al., 2007]:

$$C_P/R^* \sim A + BT + CT^2 \qquad (5)$$

where the coefficients A, B, and C are empirically adjusted to models of the specific heat of linear molecules from their translation, rotational and vibrational modes and result in $A = 2.5223$, $B = 0.77101 \times 10^{-2} \, K^{-1}$ and $C = -0.3981 \times 10^{-5} \, K^{-2}$ [Epele et al., 2007].

The temperature dependency of the specific heat capacity leads to a more complex relation between the potential temperature of the adiabatic trajectory and the pressure level of reference (see equation 19 by Epele et al. [2007]), where the value of $\theta$ has to be computed numerically using an iterative algorithm. Nevertheless a new magnitude $\tau$ with physical dimensions of temperature can be defined which verifies:

$$C_P(T)\frac{\delta T}{T} = C_P^0 \frac{\delta \tau}{\tau}. \qquad (6)$$

This new variable allows to treat the problem in exactly the same way as in the case of the ideal perfect gas but using the new "extended" potential temperature:

$$\tilde{\tau} = \tau \left(\frac{P_0}{P}\right)^{k^0} \qquad (7)$$

where $k^0 = \frac{R^*}{C_P^0}$ with $C_P^0 = C_P(T_0)$ and $T_0 = \tau(T_0)$. Thus, the relation between $\tau$ and T is given by [Epele et al., 2007]:

$$\ln\left(\frac{\tau}{T_0}\right) = \frac{A}{C_P^0} \ln\left\{\frac{T}{T_0} \exp\left[\frac{B}{A}(T - T_0) + \frac{C}{2A}(T^2 - T_0^2)\right]\right\}. \qquad (8)$$

In the current work, we use the reference values $P_0 = 1$ bar and $T_0 = 350$ K that correspond to an altitude of ~50 km in the Venusian atmosphere [Seiff et al., 1985].

According to the general definition of the potential vorticity, $\theta$ in equation (3) could be any conserved scalar quantity that (for a non-barotropic fluid) is a function of density and pressure only. Therefore, the "extended" potential temperature $\tilde{\tau}$ defined by Epele et al. [2007] is a valid alternative to potential temperature. Since the difference between $\theta$ and $\tilde{\tau}$ can be larger than the estimated error at the upper limit of the isentropic surface over which we calculate the EPV at the upper cloud level (see below), we use $\tilde{\tau}$ instead of $\theta$. Thus equation (3) becomes:

$$q \sim \zeta_{\tilde{\tau}}\left(-g\frac{\partial \tilde{\tau}}{\partial P}\right). \qquad (9)$$

## 3.2. EPV at the upper and lower clouds

The horizontal spatial structure of $q(x, y)$ depends on the wind velocity field that determines $\zeta_{\tilde{\tau}}(x, y)$, and on the three-dimensional temperature structure through $\frac{\partial \tilde{\tau}}{\partial P}(x, y)$. Thus, from our previous analyses, we can compute maps of the instantaneous distribution of $q(x, y)$ for the upper cloud layer. Temperature retrievals from VIRTIS infrared spectra (summarized in section 2.2) do not allow us to obtain temperatures below 55 km altitude in the Venus atmosphere. Therefore, we cannot derive information about the thermal distribution at the lower cloud level (at about 43 km altitude) in the same way as at the upper cloud's level. However, the $\frac{\partial \tilde{\tau}}{\partial P}(x, y)$ term can be estimated from measurements of the atmosphere's static stability, defined as $S_T = \left(\frac{dT}{dz} - \Gamma\right)$ with $\Gamma = -g/C_P$ being the adiabatic lapse rate:

$$\frac{d\theta}{dP} = -\frac{\theta}{T} S_T \frac{1}{\rho g} \tag{10}$$

and where hydrostatic equilibrium is assumed [Sánchez-Lavega, 2011]. Definitions of $\tilde{\tau}$ and $\theta$ in equations (4 – 6) and direct numerical comparison result in

$$\frac{d\theta}{dP} \approx \frac{\partial \tilde{\tau}}{\partial P}, \tag{11}$$

so, we can use equation (10) and an approximate evaluation of the static stability at the lower cloud layer (that will be presented in section 4.2.) to estimate the $\frac{\partial \tilde{\tau}}{\partial P}(x, y)$ term at the lower cloud's level.

## 4. Results

### 4.1. Upper cloud level

#### 4.1.1 Extended potential temperature from 55 to 85 km

Figure 2 shows zonally averaged results of the extended potential temperature ($\tilde{\tau}$) calculated from the thermal fields for the three dates analyzed. The latitudinal and vertical structure of $\tilde{\tau}$ is very similar in the three cases. $\tilde{\tau}$ increases with altitude showing a statically stable atmosphere, at least at the altitude range 55 – 85 km (1 – 360 mbar). $\tilde{\tau}$ decreases towards the south pole, this effect being stronger below ~66 km (~50 mbar). The results above 60 km altitude agree with the analysis by Piccialli et al. [2012] who used the usual potential temperature $\theta$ instead of $\tilde{\tau}$. At this altitude range, the polar atmosphere is slightly vertically depressed with the largest decrease of altitudes for isentropic surfaces ($\tilde{\tau}$ = constant) between 70°S and 80°S (the poleward limit of the cold collar is usually observed in this latitude range). This sinking of the isentropic surfaces with altitude towards the pole is not so strong as the one that occurs with the isobaric surfaces (P = constant) [Piccialli et al., 2012] or with the sinking of cloud top altitude [Garate-Lopez et al., 2015]. However, below 60 km, the isentropic surfaces seem to increase with altitude as shown by Piccialli et al. [2012]. The remarkable "cold collar", located typically between 60°S and 75°S at 62 km altitude in the averaged T(z, φ) maps [Garate-Lopez et al., 2015], disappears in the averaged τ̃(z, φ) maps because of the combined dependence of temperature and pressure with altitude and latitude.

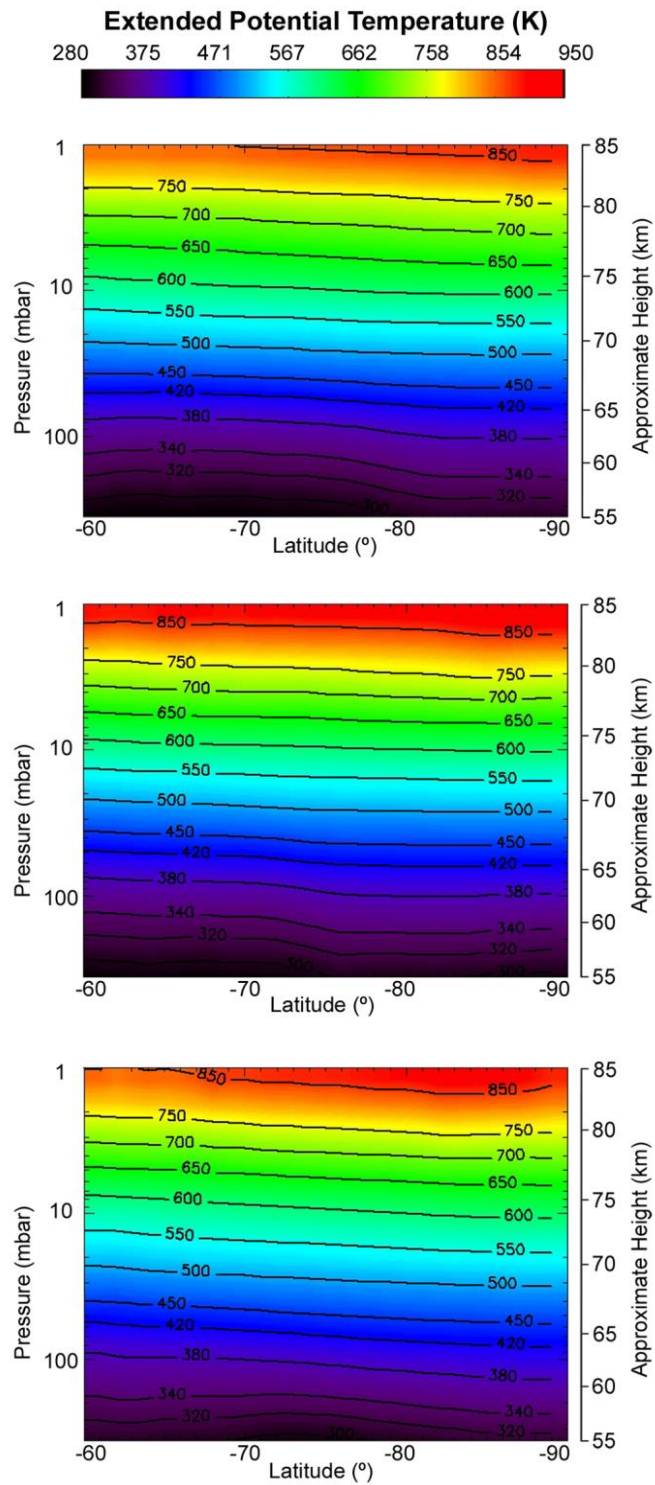

**Figure 2:** Zonally averaged extended potential temperature between 55 and 85 km altitude, on orbits 038 (top), 310 (middle) and 475 (bottom). Data was averaged in 1° bins.

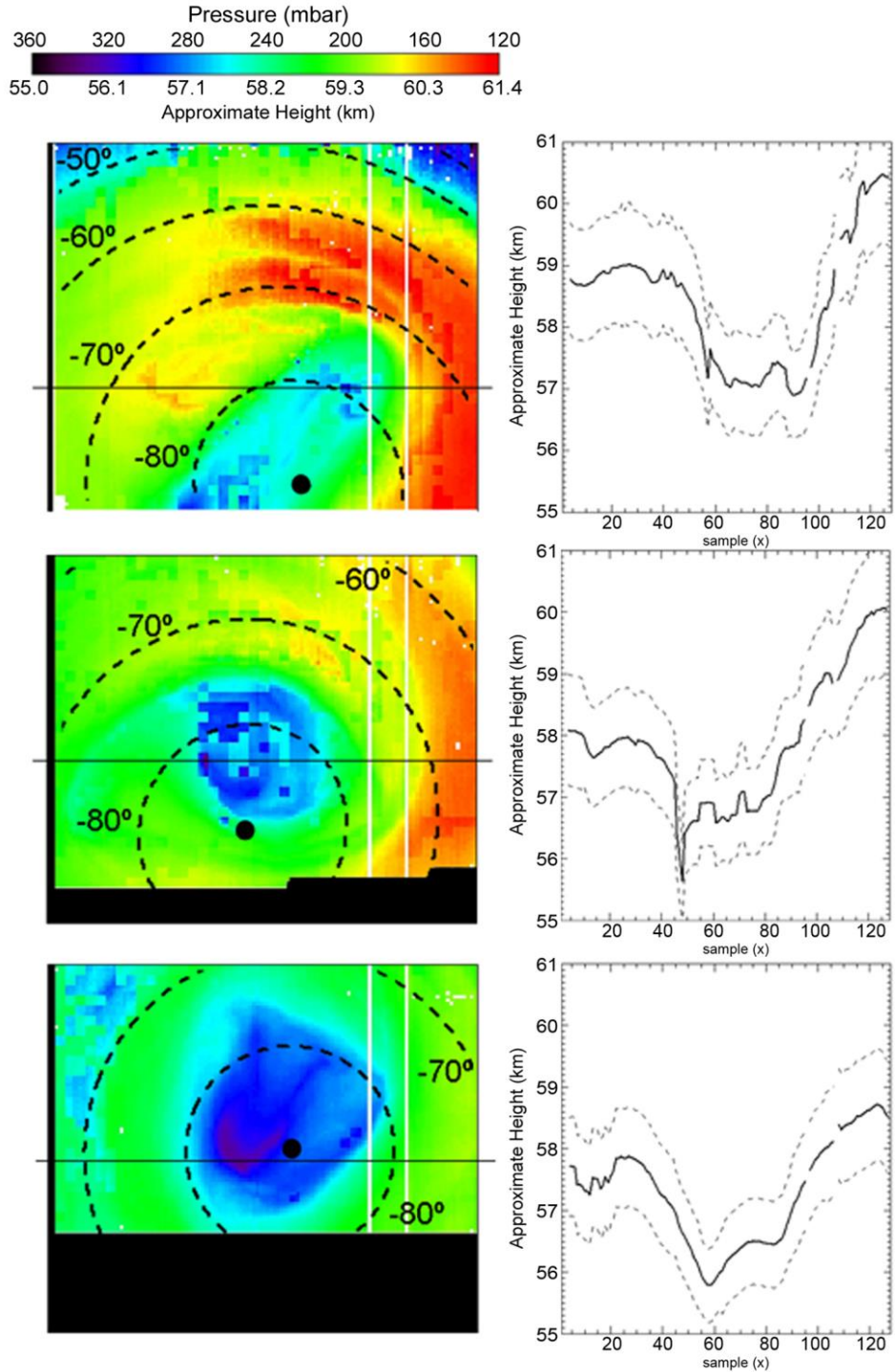

**Figure 3:** *Left*: altitude variation of the 330 K isentropic surface (with respect to the original geometry of the observations) on orbits 038 (top), 310 (middle) and 475 (bottom). The white pixels and vertical lines are due to the lack of thermal information. A certain degree of pixilation of these maps come from the noise present in the maps and depends on the cloud model parameters used in the temperature retrieval [Garate-Lopez et al., 2015]. *Right*: altitude variation over the solid lines displayed in the left columns. Dashed lines depict the altitude uncertainty range. The spatial resolution is 1 pixel ~16 km.

Figure 3 displays the altitude variation of the 330 K isentropic surface for the three dates. This is the deepest isentropic surface that can be continuously computed for the three orbits in all the pixels of the VIRTIS images. The white pixels and lines are due to the lack of thermal information and the rectilinear structures visible in the images due to the spatial resolution of the retrieval of the cloud parameters as discussed in Garate-Lopez et al. [2015]. Results for a horizontal single line in the original images are also shown for comparison. The overall shape of the vortex at the cloud top is clearly distinguishable in these isentropic altitude images. The vortex constitutes a depressed area where bright narrow features in the 3.8 or 5.1 µm images (see Figure 1) are located slightly deeper in the atmosphere. The best results in terms of less noise and spatial resolution are obtained in orbit 475, where the 330 K isentropic surface reaches the 360 mbar pressure in a region that corresponds to the local maxima of brightness emission. This pressure corresponds to an altitude of ~55 km. In the three orbits, the pressure along the 330 K isentropic surface varies by about 160 – 180 mbar, which corresponds to altitude differences of about 4 – 5 km at these atmospheric levels, comparable to the vertical scale height at these altitudes (~5.6 km). Plots on the right of Figure 3 show altitude variations of 2 – 3 km over horizontal distances of 240 – 300 km (from regions out of the warm vortex to its center) and resolve some degree of spatial structure within the vortex.

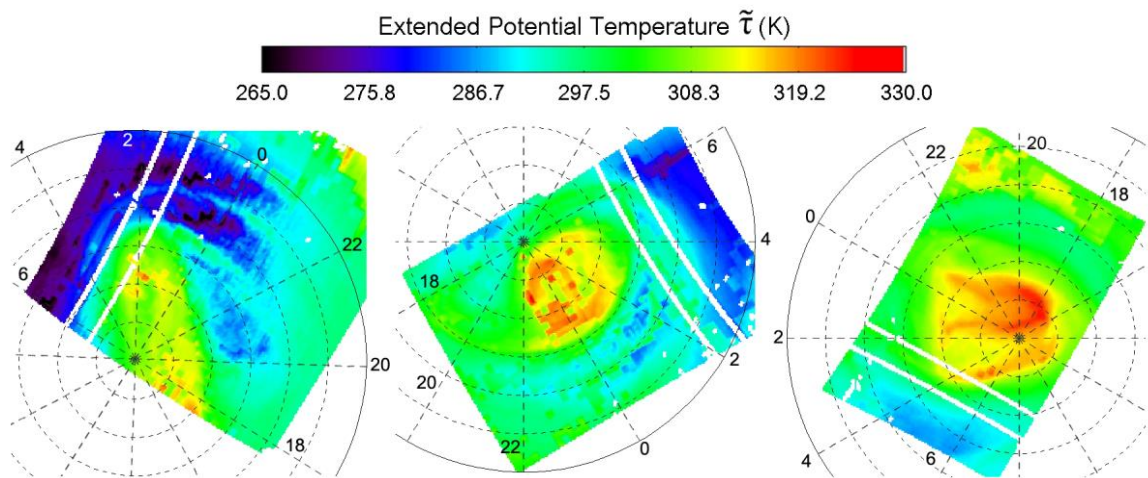

**Figure 4:** Polar maps displaying the extended potential temperature distribution at ~360 mbar (~55 km) on orbits 038 (left), 310 (center) and 475 (right). The estimated error at this pressure level is ~33 K. Latitude circles are plotted at 5° intervals from the south pole. Labels correspond to local time in hours. The white pixels and lines are due to the lack of thermal information and the presence of large square pixels present in the maps depend on the cloud model parameters used in the temperature retrieval [Garate-Lopez et al., 2015].

The horizontal distribution of $\tilde{\tau}$ at the deepest level of the thermal analysis (360 mbar, ~55 km) is shown in Figure 4. The small-scale vortex's structure characteristic of the original images (Figure 1) is also present here. Just as in the thermal structure at this vertical level, the vortex stands out as a hot region with slightly blurred warmer filaments and surrounded by colder air. Extended potential temperature differences at this pressure level can be as large as ~50 K between the cold collar and the vortex on the three orbits, agreeing with atmospheric temperature differences previously found [Garate-Lopez et al., 2015]. This means that the cold collar and the warm vortex are dynamically separated.

## 4.1.2. Vertical gradient of the extended potential temperature

The vertical gradient of the extended potential temperature, $\frac{\partial \tilde{\tau}}{\partial P}$, is the quantity that incorporates the thermal structure into the definition of EPV in equation (9). This magnitude can also be used to measure the stability of the atmosphere with respect to convection and defines those regions that are stable to vertical motions ( $\frac{\partial \tilde{\tau}}{\partial P} < 0$ ). In our previous temperature analysis [Garate-Lopez et al., 2015], we studied the static stability of the atmosphere, $S_T$ , which is equivalent to this magnitude but can be interpreted more easily. Since $S_T$ depends on vertical derivatives of temperature it was necessary to use an adequate vertical discretization that minimized errors in the vertical derivatives while preserving the information on $S_T$. We calculated the static stability for 7 atmospheric layers between 55 and 85 km considering relatively thick vertical layers that allowed to maintain the estimated static stability errors below ~10% of the adiabatic lapse rate ($\Gamma$ ~10.4 K/km on Venus's atmosphere). The thickness for each layer from top (~85 km) to bottom (~55 km) was equal to 4.0, 3.2, 3.3, 3.7, 3.2, 4.9, and 7.3 km. We here consider the same 7 atmospheric layers to compute the vertical gradient of the extended potential temperature.

The zonally averaged distribution of $\frac{\partial \tilde{\tau}}{\partial P}$ (not shown here) presents a stratified, statically stable atmosphere in this altitude range for the three dates, with absolute values increasing with altitude and no remarkable structure in any of the orbits. In contrast, the horizontal distribution does show structure, mainly in the lowermost layers. Figure 5 plots the spatial structure of $\frac{\partial \tilde{\tau}}{\partial P}$ at the two lowermost layers, which cover the altitude range reported in the literature for the upper cloud (55 – 67 km). At the 55 – 62 km altitude range the fine-scale structures show slightly smaller absolute values (less negative) than their immediate surroundings, implying that the highly variable structures present within the vortex and seen as bright at ~5 µm images are slightly less stable than any other region in the vortex. This is better appreciated in the case of orbit 475 where the filament inside the vortex is completely recovered in the $\frac{\partial \tilde{\tau}}{\partial P}$ map. At 62 – 67 km layer, the vortex is clearly distinguishable with a smaller absolute gradient than the area covered by the cold collar. In fact, the cold collar shows the most negative values of $\frac{\partial \tilde{\tau}}{\partial P}$ being, therefore, the most stable region in the polar area. These results are fully consistent and equivalent to those obtained in our previous analysis of $S_T$ [Garate-Lopez et al., 2015] and are here reiterated due to the role played by the $\frac{\partial \tilde{\tau}}{\partial P}$ term in the definition of EPV.

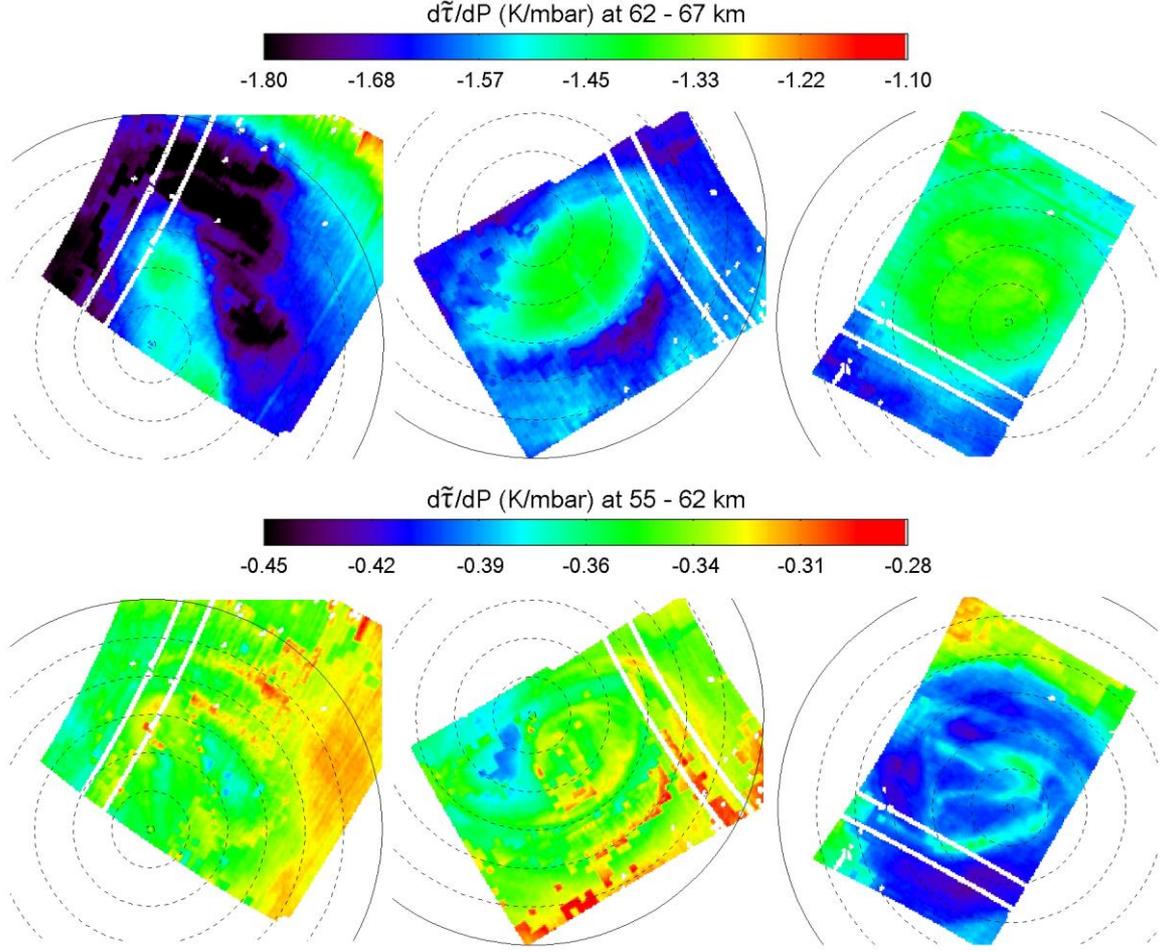

**Figure 5:** Polar maps of the vertical gradient of the extended potential temperature between 360 and 100 mbar (bottom) and between 100 and 35 mbar (top) levels on orbits 038 (left), 310 (center) and 475 (right). The estimated errors are 0.13 K/mbar and 0.14 K/mbar for the bottom and top layers, respectively. Latitude circles are plotted at 5° intervals from the south pole. The white pixels and lines are due to the lack of thermal information and the presence of large square pixels in the maps depend on the cloud model parameters used in the temperature retrieval [Garate-Lopez et al., 2015].

### 4.1.3. Ertel's potential vorticity

We assume that the winds derived over the ~5 µm images are representative of the motions at the $\tilde{\tau} = 330$ K isentropic surface which varies in the range $120 - 360$ mbar or $55 - 61.5$ km in different regions of the polar area and dates (see Figure 3). This assumption is supported by the similarity of air temperature maps at the $\tilde{\tau} = 330$ K isentropic surface (see Figure 6) with the spatial structure in the ~5 µm images (Figure 1). Even if the motions are not exactly retrieved at a constant isentropic level this assumption is still valid provided there is a small altitude difference between the isentropic surface and the vertical level where the observed motions occur or a small vertical wind shear at this altitude level as found by Hueso et al. [2015].

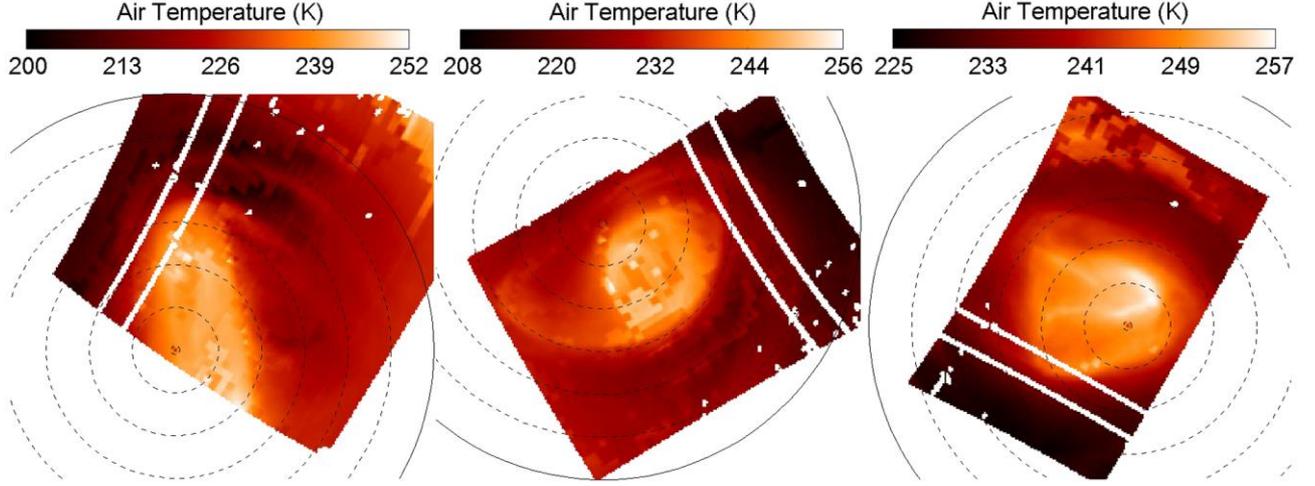

**Figure 6:** Polar projected maps of retrieved temperature over the 330 K isentropic surface on orbits 038 (left), 310 (center) and 475 (right). The estimated error is about 9 K on average. Latitude circles are plotted at 5° intervals from the south pole. The white pixels and lines are due to the lack of thermal information and the large square pixels present in the maps depend on the cloud model parameters used in the temperature retrieval [Garate-Lopez et al., 2015].

Ertel's potential vorticity (EPV) can then be calculated over the 330 K isentropic surface on the three orbits. Figure 7 shows the horizontal distribution of EPV and separates the effects from the two terms of equation (9) that define EPV (vertical component of the relative vorticity, $\zeta_{\tilde{\tau}}$ , and vertical gradient of the potential temperature multiplied by gravity, $-g\frac{\partial \tilde{\tau}}{\partial P}$). The white areas within the maps represent regions where we do not have an adequate sampling of wind measurements to retrieve the relative vorticity. We notice that the EPV does not retain the structure seen in the radiance image or in $\tilde{\tau}$ and $\frac{\partial \tilde{\tau}}{\partial P}$ maps, but it mostly resembles the distribution of the relative vorticity $\zeta_{\tilde{\tau}}$.

We found previously that peaks of relative vorticity are generally surrounded by bright features as seen in ~5 µm images [Garate-Lopez et al., 2013] and that these bright features are due to higher atmospheric temperatures [Garate-Lopez et al., 2015]. Consequently, we now find that the warmest structures are located coincident with local minima of EPV. The relation between high absolute potential vorticity values and colder temperatures is also seen on Earth's stratospheric polar vortices at large spatial-scales when the whole stratospheric vortex (that can extend from the pole to latitudes lower than 60°) is observed [Shoeberl and Hartmann, 1991].

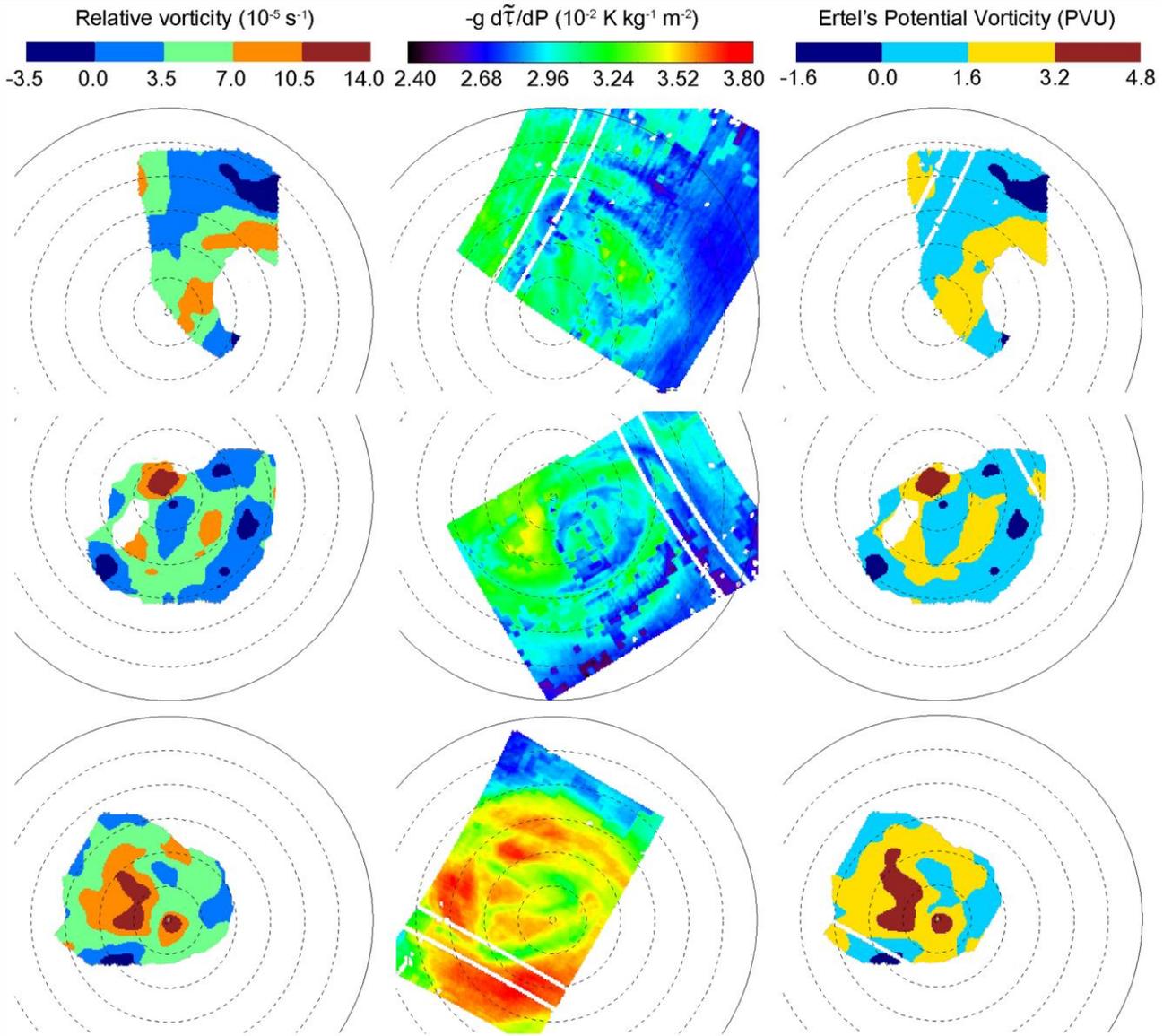

**Figure 7:** Polar maps of the vertical component of the relative vorticity (left), $-g\frac{\partial \tilde{\tau}}{\partial P}$ term (center) and potential vorticity distribution at the 330 K isentropic surface (right) on orbits 038 (top), 310 (middle) and 475 (bottom). Latitude circles are plotted at 5° intervals from the south pole. The large white areas within the maps have been eliminated due to the scarcity of wind measurements there. The white pixels and lines are due to the lack of thermal information and the large square pixels present in the maps depend on the cloud model parameters used in the temperature retrieval [Garate-Lopez et al., 2015]. Units: 1 P.V.U. = $10^{-6}$ K $\cdot$ m$^2\cdot$ kg$^{-1}\cdot$ s$^{-1}$.

The global EPV structure of the SPV of Venus at the upper cloud's level does not show any strong latitudinal gradient that could act as a mixing barrier to transported compounds. Interestingly, the structure of EPV map of orbit 310 suggests the presence of a weak ring of potential vorticity that is not related to the morphology of the vortex or to the temperature field (cold collar or warm vortex). However, the limited number of orbits analyzed hinders the assessment of the significance of this feature.

Local minima and maxima of the EPV are found close to each other with differences of up to 4 *Potential Vorticity Units* (P.V.U. $\equiv 10^{-6}$ K $\cdot$ m$^2$ $\cdot$ kg$^{-1}$ $\cdot$ s$^{-1}$). The local maxima close to the south pole on orbits 310 and 475 are probably due to numerical artifacts in the EPV derivations and should not be taken into account. Recall that we have derived EPV in the region where we have simultaneous measurements of winds and temperatures, resulting in a range of barely 20° in latitude around the pole and only on the night-side of the planet (due to constraints imposed by the thermal retrieval). Thus, it is possible, that our field of view of the SPV is not large enough to distinguish the entire high EPV area and that we only see small-scale structures within a larger vortex.

## 4.2. Lower cloud level

On many dates, we obtained cloud motion measurements for the deeper cloud at ~ 43 km observed in nighttime at 1.74 µm that are simultaneous to those of the upper cloud [Garate-Lopez et al., 2013]. However, temperature retrievals of VIRTIS spectra do not sound this deeper layer. In order to estimate the potential vorticity of the lower atmosphere, its thermal structure must be inferred from other datasets. Although temporal changes can be expected, the temperature structure of the Venus atmosphere at this altitude level in both polar regions as derived from different experiments in different epochs, seems to be quite steady [Seiff et al., 1985; Tellmann et al., 2009]. Additionally, Yamamoto and Takahashi [2015] showed results from a Venusian middle atmosphere general circulation model that predicts zonally uniform temperatures at this altitude. This is our working hypothesis for the temperature structure at this altitude level.

Vertical profiles of static stability are available from in situ measurements performed by the Pioneer Venus probes [Seiff et al., 1980] and from Pioneer Venus radio-occultation measurements [Seiff et al., 1985]. Besides, the VIRA model [Seiff et al., 1985] integrates much of the practical knowledge of the Venusian atmosphere prior to Venus Express. Profiles of static stability of the atmosphere at the four Pioneer Venus probe entry sites show a stability peak at about 43 km altitude (see Figure 17 in Seiff et al. [1980]). On the other hand, the Pioneer Venus radio-occultation data and the VIRA model present a smooth decrease of the atmosphere's static stability towards the pole [Seiff et al., 1985]. We have constructed a latitude-dependent distribution of $S_T$ at the lower cloud altitude based on a quadratic fit of the Pioneer Venus North probe that fell at ~60°N, Pioneer Venus radio-occultation results at latitudes higher than 55° and VIRA data between 60° and 90° (see Figure 8):

$$S_T = 0.0045\varphi^2 - 0.7853\varphi + 34.8912 \ . \tag{12}$$

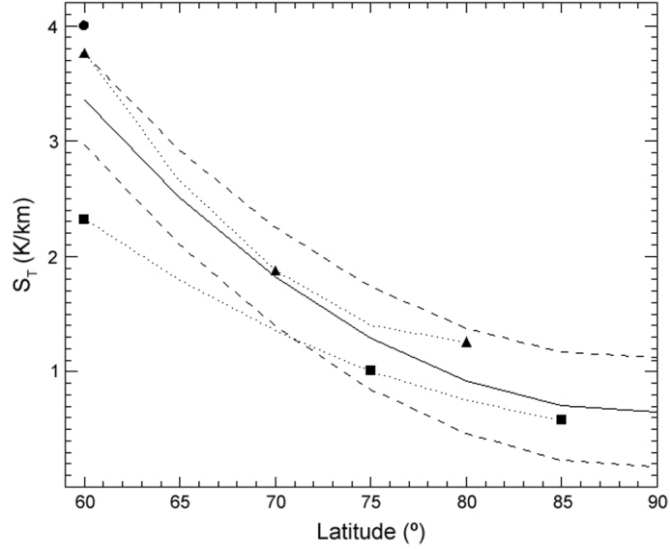

**Figure 8:** Pioneer Venus' radio-occultation (triangles) and VIRA (squares) data between 60° and 90° has been used together with the measurement of the Pioneer Venus North probe (dot) in order to describe the static stability of the atmosphere at the 43 km altitude. The continuous black line shows a two-degree fit to all data. Dashed lines show the $\pm 1\sigma$ curves.

The global stability decay towards the pole agrees with experimental results from Tellmann et al. [2009] who analyzed the VeRa radio occultation data from Venus Express and found that the stable layer below the upper cloud does not appear in all their high latitude $S_T$ profiles, meaning that static stability values close to zero are possible there. Imamura et al. [2014] used theoretical arguments to state that lower cloud convection and less stability are caused by the lower solar irradiation on the upper cloud at high latitudes.

The low values of the static stability at high latitudes and the fit to several $S_T$ data sets represented by Figure 8 means that our EPV results at the lower cloud level should be viewed critically and considered as first order assessment.

Using equations (10 – 12), the extended potential temperature definition and the same reference values of $P_0 = 1$ bar and $T_0 = 350$ K as at the upper cloud's level, it is possible to calculate the $-g \frac{\partial \bar{\tau}}{\partial P}$ term appearing in the definition of potential vorticity. The calculation is done for an average pressure level for the lower cloud. We have used P = 2.2 bar and T = 378.5 K, based on the measurements of Pioneer Venus at ~43 km altitude and between 55° and 90° [Seiff et al., 1985].

Combining this result with our previous measurements of the relative vorticity from the tracking of features seen at 1.74 µm images [Garate-Lopez et al., 2013] and using equation (9), we have tentatively estimated the EPV distribution at the lower cloud's level. Figure 9 shows the EPV maps, together with the two terms appearing in equation (9), for orbits 038 and 310. Data from orbit 475 could not be used since the VIRTIS-M-IR images at 1.74 µm were not good enough to provide cloud motion measurements. The white areas within the maps represent regions where the lack of enough wind measurements prevents to obtain meaningful results. Typical values of EPV at the lower cloud layer are on the order of $2.0 \times 10^{-2}$ P.V.U.

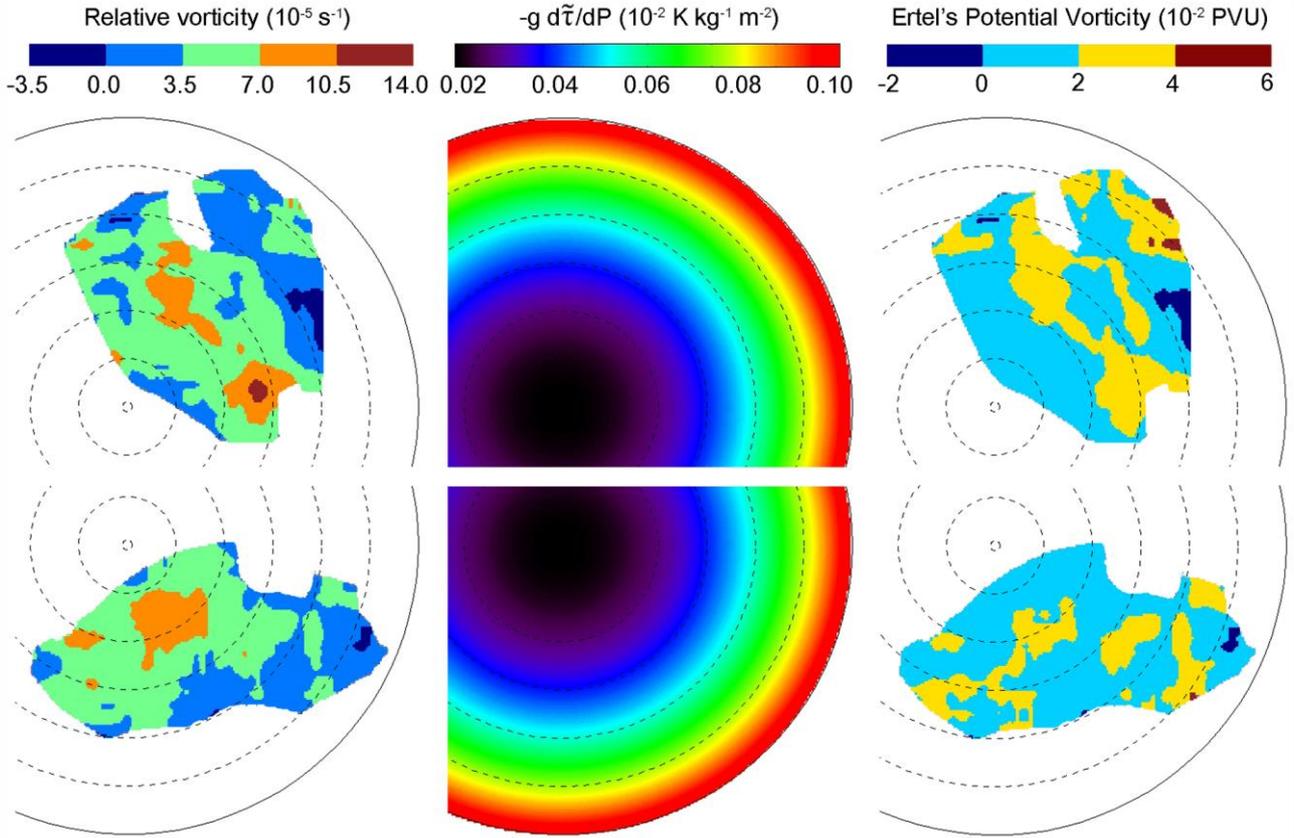

**Figure 9:** Polar maps of the vertical component of the relative vorticity (left), $-g\frac{\partial \tilde{\tau}}{\partial P}$ term (center) and potential vorticity distribution at the lower cloud's level (right) on orbits 038 (top) and 310 (bottom). Latitude circles are plotted at 5° intervals from the south pole. The large white areas within the maps have been eliminated due to the scarcity of wind measurements there. Units:
1 P.V.U. $= 10^{-6}\,\mathrm{K} \cdot \mathrm{m}^2 \cdot \mathrm{kg}^{-1} \cdot \mathrm{s}^{-1}$.

A direct comparison of EPV at both cloud layers is not immediate. There is a difference of two orders of magnitude in the potential vorticity, being higher at the upper cloud's level (Figure 7) than at the lower cloud's level (Figure 9). This is due to two effects: on the one hand, the large variation of the $-g\frac{\partial \tilde{\tau}}{\partial P}$ term, which contains high values of static stability in the upper cloud and much lower values in the lower cloud. On the other hand, there is an inverse variation with pressure in the definition of potential temperature. Note that the range at which $-g\frac{\partial \tilde{\tau}}{\partial P}$ varies at 55 – 62 km is not enough to have a significant effect on the spatial distribution of potential vorticity, as previously discussed. At the upper cloud the vorticity term dominates with respect to the thermal contribution and determines most of the appearance of potential vorticity map. Contrarily, in the lower cloud level the slowly varying values of $-g\frac{\partial\tilde{\tau}}{\partial P}$ between 75°S and the south pole tend to homogenize the distribution of potential vorticity, thereby, smoothing the structure present in the relative vorticity map and impeding us to characterize the details of the potential vorticity at this lower depth.

In order to compare the values of EPV at both cloud layers we have normalized $q(x, y)$ in each layer by the horizontal mean value of $-g\frac{\partial\tilde{\tau}}{\partial P}$ (not shown here). This technique has been used previously by Read et al. [2009] and Piccialli et al. [2012] in global estimations of EPV at different altitudes in Saturn and Venus, respectively. The normalized potential vorticity values at the upper cloud's level (from -3.6 to 15.9 x $10^{-5}$ s$^{-1}$) are 3 – 4 times those at the lower cloud level (from -0.9 to 5.2 x $10^{-5}$ s$^{-1}$) meaning that the vortex strength is higher at the upper cloud but extends towards the lower atmosphere. The spatial structure remains similar to the not normalized EPV maps.

### 4.3. Error Analysis

Combining the estimated errors of the thermal retrieval and the estimation of $3.5 \times 10^{-5}$ s$^{-1}$ error associated to the maps of the vertical component of the vorticity, the error propagation from equation (9) results in an average value of ~1.6 P.V.U. for the Ertel's potential vorticity at the upper cloud's level for the three orbits. Although there are many unknowns that complicate the EPV analysis and may enlarge the uncertainty of the results, we consider this a good global estimation of errors. The main unknowns are: (1) The analysis is restricted to the night-side of the planet (we can obtain motions in the night and day-side but temperatures are only available in the night-side). (2) The motions are obtained from the displacements of features in ~3.8 or 5 µm thermal images that in fact correspond to a range of vertical altitudes that is difficult to precisely constrain. Our calculation of EPV over an isentropic surface assumes that the vertical gradient of motions is small enough to use the winds derived over the thermal images as representative of the motions at the 330 K isentropic surface. (3) The limited vertical resolution in the thermal retrieval impedes investigating thermal variations at vertical scales lower than ~7 km in the layer between 55 and 62 km altitude (which is of the order of the gas scale height).

At the lower cloud the error propagation results in an average uncertainty of ~2 x $10^{-2}$ P.V.U., when considering the $\pm 1\sigma$ standard deviation curves of the quadratic fit of Pioneer Venus and VIRA data as the $S_T$ error (see Figure 8). At this level, we do not know the precise horizontal thermal structure and the static stability of the atmosphere is estimated from data that comes mainly from the Northern Hemisphere and previous missions (not coinciding in time with our wind analysis). We assume that the wind measured using 1.74 µm images are representative of the motions at the altitude where the static stability observed by Pioneer Venus and modeled by VIRA shows a peak below 50 km. Nevertheless, we consider these factors may have a limited impact in the EPV calculation and that the EPV errors given above are large enough to encompass variations associated to these factors.

All in all, the significance of the structures visible in EPV maps derived here can be estimated from the range of variation in EPV that is 3 times larger than the uncertainties at both cloud levels. Although this is not an optimal S/N ratio, EPV fields from VIRTIS data are likely to be the only available ones for the study of the vortex because the Akatsuki spacecraft cannot observe the polar areas from its equatorial orbit.

### 4.4. Short-term evolution

In some of the Venus Express orbits, several high-resolution observations of the vortex were obtained every 15 minutes providing an excellent data set for short-term dynamics. We have extensively analyzed VEX orbit 475 grouping the data in 6 image pairs separated by 1 hour. The total time covered is ~ 6 hours (about 1/9th of the rotation period of the vortex). Each image pair has been used to obtain accurate wind measurements and each VIRTIS qube to retrieve thermal profiles (as explained in section 2).

Figure 10 depicts the retrieved atmospheric temperatures and the term $-g\frac{\partial \tilde{\tau}}{\partial P}$ over the 330 K isentropic surface for six VIRTIS data qubes on orbit 475. Since the fine-scale features seen in 5 μm radiance images are recovered in the temperature maps, radiance images are not shown. Some bins in boxes of 6 x 6 pixels, as well as a few white bins, appear in the data. The former is due to the cloud parameters' analysis and the latter because the temperature retrieval did not provide good results there [Garate-Lopez et al., 2015]. These polar projections show a counterclockwise cyclonic rotation of the vortex with only minor changes in the fine-scale filamentary structure inside the vortex.

Figure 11 shows the short-term evolution of the vertical component of the relative vorticity field at the upper cloud's level and of Ertel's potential vorticity distribution over the $\tilde{\tau} = 330$ K isentropic surface. Both variables show essentially the same spatial variation. Local minima and maxima, which apparently extend over time, are found in all the six qubes, but we do not see any clear rotation of the structures appearing in the EPV maps as we do in the evolution of the temperature (and radiance) field. This result is surprising since the EPV is expected to be a conserved quantity and, therefore, a tracer of fluid motions for atmospheric flows. Taking into account the unpredictable and highly variable nature of the vortex, we could speculate about possible sinks and sources of potential vorticity at the polar region of Venus's atmosphere, but the lack of correlation between the behavior of the thermal and EPV structures is more likely related to the large errors (about 1/3rd of the EPV) and the reduced spatial resolution of about 525 km (due to the spatial derivatives in the relative vorticity) in our analysis. Interestingly, the anti-correlation between warm features at 360 mbar (~55 km) and high values of EPV remains.

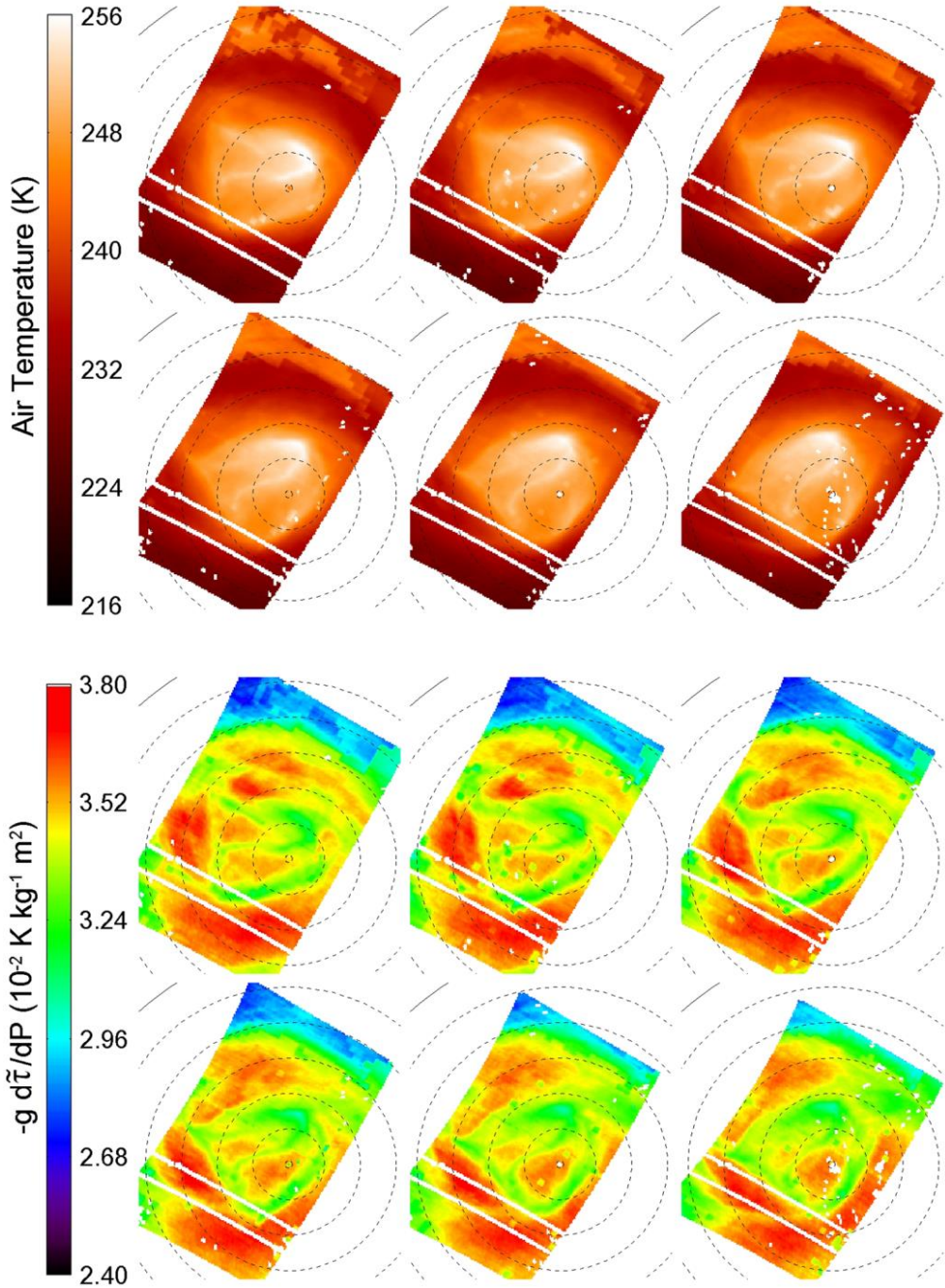

**Figure 10:** Polar maps showing the short-term evolution of the atmospheric temperature (top two rows) and $-g\frac{\partial \tilde{\tau}}{\partial P}$ (bottom two rows) fields over the 330K isentropic surface during orbit 475. From left to right and top to bottom, data qubes correspond to VI0475_04, VI0475_08, VI0475_12, VI0475_16, VI0475_20 and VI0475_24, and are separated by intervals of ~60 minutes with a total time of ~ 6 hours (~1/9th of the rotation period of the vortex). The white pixels and lines are due to the lack of thermal information and the large square pixels present in the maps depend on the cloud model parameters used in the temperature retrieval [Garate-Lopez et al., 2015].

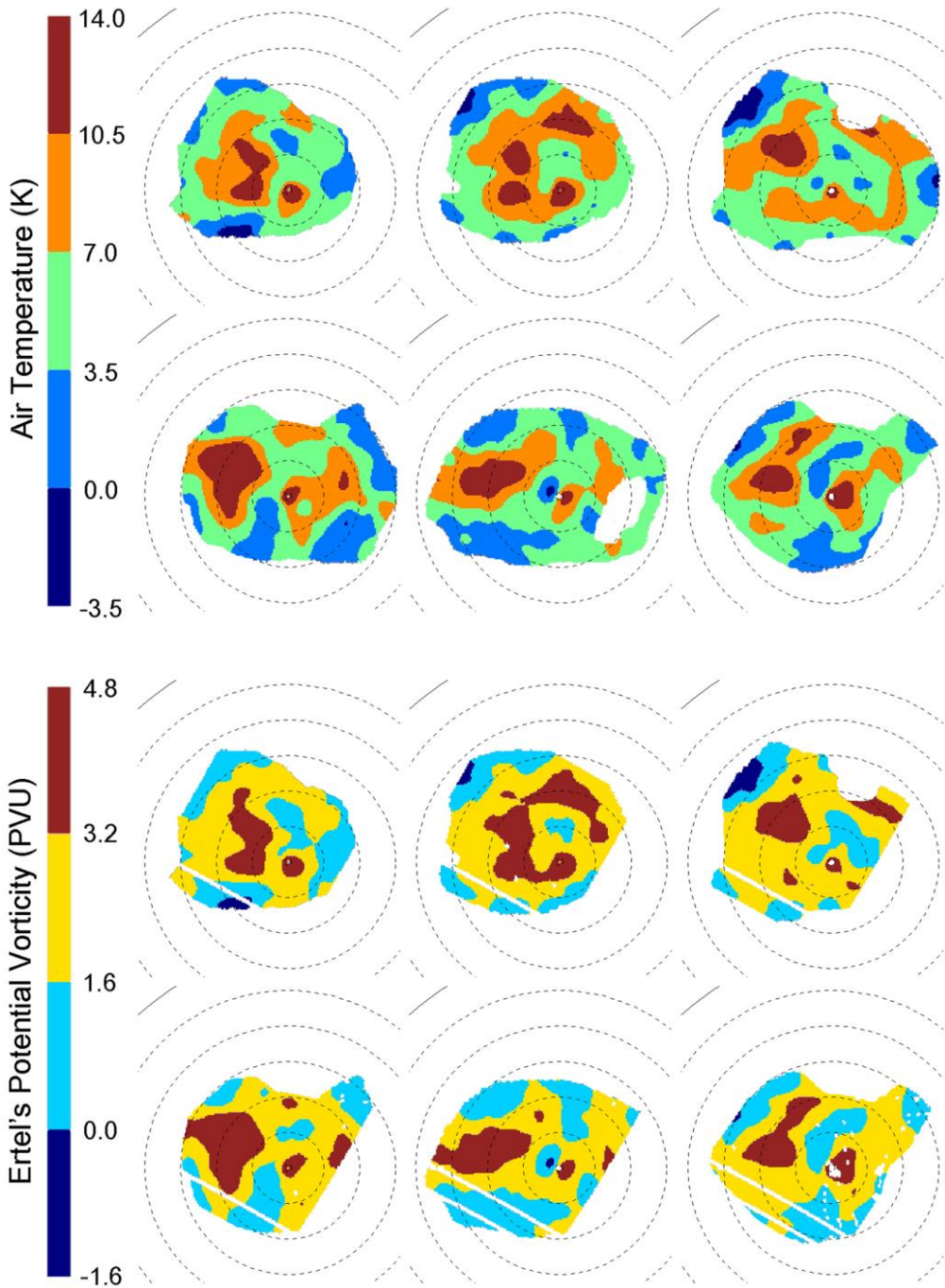

**Figure 11:** Polar maps showing the short-term evolution of the vertical component of the relative vorticity field (top two rows) and Ertel's potential vorticity (bottom two rows) distribution over the 330 K isentropic surface during orbit 475. From left to right and top to bottom, data qubes correspond to VI0475_04, VI0475_08, VI0475_12, VI0475_16, VI0475_20 and VI0475_24, and are separated by intervals of ~60 minutes with a total time of ~ 6 hours (~1/9[th] of the rotation period of the vortex). The large white areas within the maps have been eliminated due to the scarcity of wind measurements there. The white pixels and lines are due to the lack of thermal information and the large square pixels present in the maps depend on the cloud model parameters used in the temperature retrieval [Garate-Lopez et al., 2015].

The mean structure observed in the EPV maps in this sequence points to a weak ring of potential vorticity which is particularly clear in the second map in the sequence (VI0475_08 data qube) or when averaging all six panels. This structure is also present in the EPV map corresponding to orbit 310 at the upper cloud's level (Figure 7). This feature, the weak vorticity ring, not centered in the pole, appears in approximately half of the EPV maps of the upper cloud. However, the low number of orbits analyzed and the limited spatial coverage impede us to infer conclusions about the significance of this ring. If confirmed, the ringed structure of vorticity would be a trait in common with Mars' polar vortices [Mitchell et al., 2014], while the extended structure of the vortex along large vertical scales (it extends vertically at least 20 km) is similar to Earth's polar vortices. Hence, Venus' SPV apparently shares features with Mars' and Earth's polar vortices and may therefore be an intermediate case between both atmospheres.

## 5. Discussion

The SPV has been continuously observed during the VEX mission (from 2006 till 2008 by VIRTIS and VMC instruments and from 2008 till 2014 by VMC showing only the external part of the vortex), so it can be considered as a long-lived feature of Venus' atmospheric dynamics. Because of the hemispheric symmetry of the general circulation of Venus at cloud level [Limaye, 2007; Peralta et al., 2007], and the similarities of thermal structures found by the Venus Express VeRa and VIRTIS-H instruments on both hemispheres (Tellmann et al., 2009; Migliorini et al., 2012), we conjecture that the same conclusion is valid for the northern polar vortex observed in the Mariner 10 and Pioneer-Venus epochs. This is in contrast to Earth and Mars vortices that show an annual trend of formation and destruction linked to the seasonal insolation cycle [Schoeberl & Hartmann, 1991; Kieffer et al., 2000]. The vortices on these two planets are related to their surface structure and icy polar caps, which is not the case of Venus. Venus polar vortices are embedded within the clouds, at about ~10 scale heights above the surface, whereas Earth's polar vortex is located at ~2 scale heights from the surface and essentially no clouds form on them. On Venus heat deposition at the polar cloud level is probably a basic ingredient in driving the formation of the warm polar vortices, a mechanism that doesn't play any role on Earth and Mars. Apparently, Venus' polar vortices are free, high altitude atmospheric features, not linked to Venus' polar topography. In all three planets, the vortices extend meridionally from the pole to latitudes close to ~60° [Garate-Lopez et al., 2013; Mitchell et al., 2014], or down to even lower latitudes in the case of the Earth [Waugh & Polvani, 2010].

On Earth, the polar vortices extend vertically from the tropopause (~10 km) up to ~50 km [Shoeberl & Hartmann, 1991], but they intensify their circulation at ~35 km in the stratosphere (in particular during winter time over Antarctica). Earth's polar vortices are highly variable with velocities up to 90 m/s at the stratospheric polar jet [Schoeberl & Hartmann, 1991; Waugh & Polvani, 2010]. On Mars, polar vortices have been traced at about 40 km altitude by Mitchell et al. [2014], who found that the mean strength of the jet maximum is ~70 m/s in the Southern Hemisphere winter and ~130 m/s in the Northern Hemisphere winter. On Venus, the polar vortices have been observed to exist from ~43 km to 80 km, intensifying at the upper cloud level about 63 km altitude [Garate-Lopez et al., 2015]. Contrarily to what happens on Earth and Mars, Venus's SPV does not seem to be related to a polar jet since we did not see any localized jet in the instantaneous [Garate-Lopez et al., 2013] or mean wind fields [Hueso et al., 2015].

The comparison of polar maps of the EPV in the three planets shows similarities and differences. On Venus and Earth, the EPV field shows typically concentration of vorticity patches with maximum values of about 5 P.V.U. at 55 – 62 km on Venus (Figure 7) and 1200 P.V.U. at ~35 km on Earth [Clough et al., 1985]. Zonal wavenumbers 1 and 2 are typically observed on Earth but on Venus more variability is found. On both planets the EPV is of the same order of magnitude (0 – 10 P.V.U.) close to the tropopause (at ~60 km on Venus, see Figure 7, and at ~10 km on Earth [Kunz et al., 2011]). The EPV decrease by two orders of magnitude in approximately 20 km altitude variation is also a common characteristic of Venus' (Figures 7 and 9) and Earth's polar vortices [Clough et al., 1985; Kunz et al., 2011; Harvey et al., 2009].

On Mars, the vortex has a ring-like structure with EPV values of ~$10^3$ – $10^4$ P.V.U. at ~35 km altitude [Mitchell et al., 2014; Montabone et al., 2014]. Some of the Venusian EPV maps presented in this paper hint at a weak ringed structure around the vortex in the upper cloud region. We do not find a correlation between the ringed structure and the vortex morphology or temperature field (cold collar or warm vortex). On orbits 310 and 475 the weak ring is observed in those regions where the zonal wind decreases more rapidly toward the pole (in agreement with the second term of the equation 1). On orbit 038 the zonal wind shows a similar decrease toward the pole, but the vorticity ring is not seen in the EPV maps. Hence, the weak vorticity ring seems to grow from the combination of the three terms in equation (1). Importantly, we must recall that we have analyzed a limited number of dates and that the EPV maps have been derived in the region where we have simultaneous measurements of winds and temperatures resulting in a range of barely 20° in latitude around the pole and only on the night-side of the planet. Thus, it is possible that our limited perspective of the south polar region impedes us from distinguishing the entire high EPV area and that we only see small-scale structures within a larger vortex.

On Earth, if we compare the shape of the polar vortex (defined by an ellipse covering the whole vortex), we see that it rotates with height in the Northern Hemisphere [Mitchell et al., 2014]. On Mars, the orientation of the polar vortex can change strongly from one day to another during certain times of the season, but it remains remarkably coherent with height at all times. However, the horizontal projection of the Martian vortices decreases dramatically with height. In terms of cloud morphology, the SPV of Venus preserves its global shape (oval, circular or irregular shape seen as a bright region at the upper cloud level) throughout 20 km in the vertical but the fine-scale structure is different at the lower and upper cloud levels (see Figure 1). However, it is difficult to study the vertical coherence of its EPV distribution since the area where motions can be retrieved at the upper and lower cloud levels is not exactly the same, we cannot retrieve the exact thermal field in the lower cloud level, and because of the limited number of orbits analyzed.

The comparison between the three planets shows that the Venus polar vortices are weaker in terms of the EPV by 3 to 4 orders of magnitude relative to Earth and Mars. However, they are permanent, which is not the case for Earth and Mars. This reflects the differences between the three planets in the ingredients involved in the recipe of their formation: planetary rotation (fast and similar on Earth and Mars; slow in Venus), seasonality (fundamental in Earth and Mars; lacking on Venus), role played by the surface (fundamental in Earth and probably Mars; not apparent on Venus), and heat deposition on clouds (essential for Venus; absent in Earth and Mars). Titan's polar vortex [Teanby et al., 2008; de Kok et al., 2014; West et al., 2015] could represent an interesting case for future studies since the polar clouds show the emergence of a seasonal cloud similar in shape to Venus SPV. Additionally, Titan's intermediate rotation rate between Venus and Earth-Mars and the presence of strong seasonal effects like Earth and Mars add intriguing properties to the polar vortices puzzle.

So far, there have been few attempts to model the structure and dynamics of Venus' polar vortices. On the one hand, Limaye et al. [2009] tried to simulate the S-shape feature observed in VIRTIS images during the early stage of the Venus Expres mission by means of a two-dimensional non-linear, non-divergent barotropic model. On the other hand, Lee et al. [2010] investigated the cloud structures produced by the circulation and eddy transport by implementing a passive cloud condensation scheme into a Venus General Circulation Model with a superrotating middle atmosphere. However, these two works fail to reproduce the highly variable morphology of the vortex. Yamamoto and Takahashi [2015] have investigated the polar vortex in the presence of a thermal tide with a Venus middle atmosphere general circulation model. In that model the warm polar air mass at the cloud top is maintained by the thermal wind associated with a high latitude jet and the cold collar is enhanced by a polar diurnal tide. Interestingly, the geometric centre of their warm oval is displaced from the pole about 10° by the diurnal tide, and transient dipole and tripole structures appear as a consequence of the superposition of a transient baroclinic wave and a diurnal thermal tide. However the model implied thermal tides that are stronger than those so far measured in Venus polar atmosphere [Peralta et al., 2012].

To date no study has established a link between the cloud morphology of the Venus' polar vortex, its motions and its relation with the overall atmospheric dynamics. Therefore, extending the study of long- and short-term evolution of the Ertel's potential vorticity will help understand the dynamics of the vortex. Further studies would also include the calculation of EPV at higher altitude levels by estimating the thermal winds by an appropriate cyclostrophic wind equation for polar latitudes.

## 6. Summary

Venus' South Polar Vortex (SPV) is a long-lived, highly variable structure of Venus' atmosphere subject to strong changes and erratic motions around the south pole [Piccioni et al., 2007; Luz et al., 2011; Garate-Lopez et al., 2013]. This is related to the fact that the motions defining the vortex are weak when compared to the environmental winds in its surrounding region. From the point of view of relative vorticity the vortex is a weak cyclone. However, the thermal imprint of the vortex is strong with large differences of temperature at the same pressure level.

We summarize the main conclusions from this study in the following:

- The vortex is a highly vertically depressed structure when observed in isentropic surfaces from 55 to 85 km, at least in the three dates analyzed here. The 330 K isentropic surface (the deepest we have access to in all the pixels of the VIRTIS images) varies from 62 km altitude at the cold collar region to 55 km inside the vortex. The vortex itself experiences a strong altitude variation of 2 – 3 km in horizontal distances of 240 – 300 km between regions out of the warm vortex and its center over the 330 K isentropic surface. This is most probably related to the global atmospheric circulation formed by a meridional Hadley cell that transports air from higher altitudes downwards at the polar region, heating the air and forming the vertically depressed structure [Read, 2013].

- Ertel's potential vorticity's horizontal distribution at the upper cloud's level does not retain the structure seen in the radiance image or in the temperature maps, but resembles the distribution of the relative vorticity. The kinetic component dominates with respect to the thermal structure at the upper cloud's level, while in the lower cloud's level the low $-g\frac{\partial \bar{T}}{\partial P}$ values tend to homogenize the EPV distribution between 75° and 90°S respect to $\zeta_{\bar{T}}$ distribution.

- The warm highly variable fine-scale features seen in the ~5 µm images that define the SPV of Venus are located in regions where the EPV over the 330 K isentropic surface is locally minimum. This is remarkable since the relation between high absolute potential vorticity values and colder temperatures is also seen on Earth's stratospheric polar vortices.

- Although a clear rotation of the general EPV distribution is not appreciated in the short-term evolution, as it is in radiance and temperature maps, the anti-correlation between warm features at 55 – 62 km and high values of EPV over the 330 K isentropic surface (located at the same altitude range) remains.

- The structure observed in many of the EPV maps at the upper cloud's level point to a weak ring of potential vorticity without any strong latitudinal gradient of EPV, as should be expected in the presence of a mixing barrier. However, the limited number of orbits analyzed does not allow us to draw a clear conclusion about the significance of this feature. Nevertheless, local minima and maxima of EPV are found close to each other with differences of up to ~4.0 ± 1.6 P.V.U.

- Values of EPV at the lower cloud are only temptative and represent the mean EPV at this layer ($2 \times 10^{-2}$ P.V.U.) and the range of variation expected due to the vortex ($2 \times 10^{-2}$ P.V.U.).

**Acknowledgments**

The data for this paper is available at ESA's Planetary Science Archive in Venus Express / VIRTIS instrument's dataset (ftp://psa.esac.esa.int/pub/mirror/VENUS-EXPRESS/VIRTIS/). The data supporting the figures could also be requested from Itziar Garate Lopez (itziar.garate@ehu.eus). We wish to thank ESA for supporting the Venus Express mission, ASI (by the contract I/050/10/0), CNES and the other national space agencies supporting the VIRTIS instrument onboard Venus Express and their principal investigators G. Piccioni and P. Drossart. This work was supported by the Spanish project AYA2012-36666 with FEDER support, Grupos Gobierno Vasco IT-765-13 and by Universidad País Vasco UPV/EHU through program UFI11/55.